\documentclass[a4paper]{jpconf}
\usepackage{graphicx}
\usepackage{float}
\usepackage{subcaption}
\usepackage{textcomp}
\usepackage{wrapfig}
\def\aa{\textrm{a\kern -1.3ex\raisebox{0.6ex}{$^\circ$}}}

\begin{document}

\title{The HIBEAM/NNBAR Calorimeter Prototype}

\author{K Dunne$^1$,
        B Meirose$^{1,3}$, 
        D Milstead$^1$, 
        A Oskarsson$^3$, \\
        V Santoro$^2$, 
        S Silverstein$^1$ and 
        S-C Yiu$^1$}
        
\address{$^1$ Department of Physics, Stockholm University, 106 91, Stockholm, Sweden}
\address{$^2$ European Spallation Source ERIC, 225 92, Lund, Sweden}
\address{$^3$ Fysiska institutionen, Lunds universitet, 221 00, Lund, Sweden}

\ead{katherine.dunne@fysik.su.se}

\begin{abstract}
The HIBEAM/NNBAR experiment is a free-neutron search for $n \rightarrow$ sterile $n$ and $n \rightarrow \bar{n}$ oscillations planned to be installed at the European Spallation Source under construction in Lund, Sweden. A key component in the experiment is the detector to identify $n-\bar{n}$ annihilation events, which will produce on average four pions with a final state invariant mass of two nucleons, around $1.9$\,GeV. The beamline and experiment are shielded from magnetic fields which would suppress $n \rightarrow \bar{n}$ transitions, thus no momentum measurement will be possible. Additionally, calorimetry for particles with kinetic energies below 600\,MeV is challenging, as traditional sampling calorimeters used in HEP would suffer from poor shower statistics. A design study is underway to use a novel approach of a hadronic range measurement in multiple plastic scintillator layers, followed by EM calorimetery with lead glass. A prototype calorimeter system is being built, and will eventually be installed at an ESS test beam line for \textit{in situ} neutron background studies.
\end{abstract}

\section{Introduction}
The HIBEAM/NNBAR experiment~\cite{White} is a future multi-stage experiment designed to search for baryon number violating processes with free-neutron beams at the European Spallation Source (ESS). These processes include transitions of neutrons to sterile neutrons and antineutrons. This paper focuses on the detector needed to observe the spontaneous transitions of neutrons to antineutrons via the annihilation of an antineutron with a nucleon in a carbon target which will produce an average of 4--5 pions with a final state invariant mass near $1.9$\,GeV~\cite{BLG}. The experiment will deliver, in its second stage (termed NNBAR) an ultimate expected sensitivity three orders of magnitude greater than the previous free-neutron search at the Institut Laue-Langevin (ILL)~\cite{Baldo}. The sensitivity improvement comes from a combination of three decades of improvements in neutronics, detector technology and background rejection techniques, and a dedicated high flux neutron beam line at the ESS. The earlier stage of the experiment (HIBEAM) will use a lower flux beamline for sterile neutron searches and for neutrons converting to antineutrons via sterile neutron states with a sensitivity increase in oscillation time of up to an order of magnitude. For both stages, an annihilation detector is needed with similar features. Thus while the detector prototype described in this paper is discussed in the context of the later NNBAR stage, it should also be taken to be relevant for the HIBEAM stage.  

\begin{figure}
\begin{center}
\includegraphics[width=.85\linewidth]{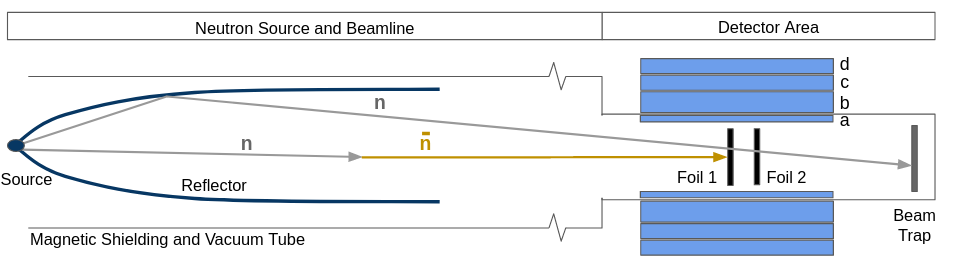}
\setlength{\belowcaptionskip}{-12pt}
\caption{A schematic representation of the NNBAR experiment (not to scale). The detector subsystems are (a) silicon tracking detector (b) time projection chamber (c) Hadronic Range Detector (d) Lead glass EM Calorimeter}
\label{fig:nnbar-det}
\end{center}
\end{figure}

The full NNBAR experiment will use a dedicated beam port at the ESS that will provide access to upper and bottom moderators~\cite{santoro}. A schematic overview of the NNBAR experiment is shown in Figure~\ref{fig:nnbar-det}. The NNBAR detector will reconstruct multi-pion antineutron annihilation events in a thin carbon foil target, while using using kinematic and timing characteristics for background rejection. 
The main detector subsystems are a silicon based tracking detector between the target foil and the vacuum tube to aid vertex reconstruction. Following the vacuum tube is a time projection chamber (TPC) for track information and particle identification. It is anticipated that silicon strip detectors will improve the vertex reconstruction by providing two space points within the tube coupled to track information from the TPC outside the vacuum. Following the TPC is the calorimeter system which will be used for triggering on annihilation events and energy and position reconstruction. The baseline proposal is based on a range measurement for hadronic particles using scintillating plastic layers followed by a EM calorimeter composed of lead glass blocks which measure energy deposits through Cherenkov radiation. 
\begin{wrapfigure}[26]{r}{5cm}
\includegraphics[width=5cm]{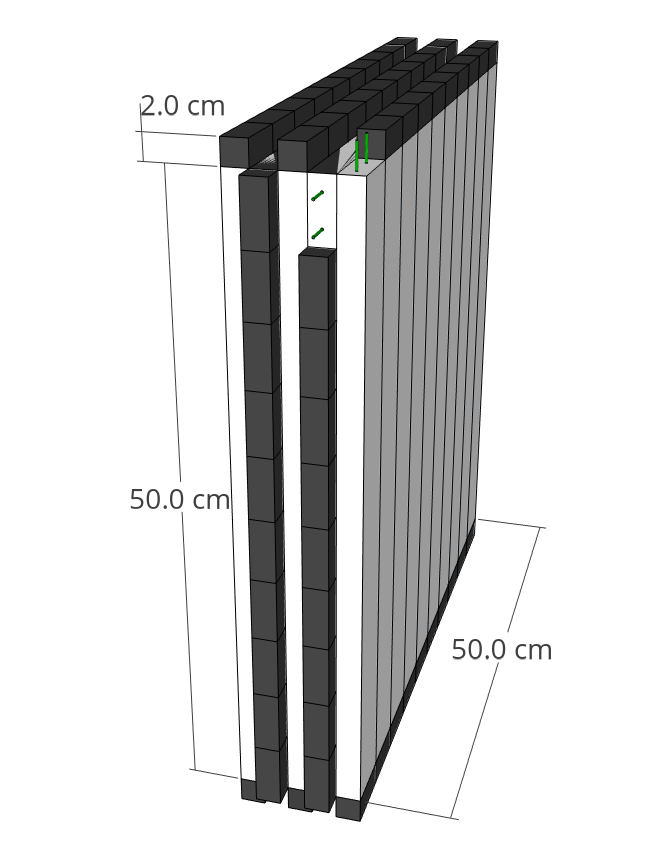}
\caption{A CAD drawing of the Hadronic Range Detector, consisting of five layers of $5.18\times2\times50$\,cm$^{3}$ staves arranged with alternating layers perpendicular to one another. The black boxes on the edge of he staves represent the electronics enclosure. The front enclosures are removed to show the arrangement of the WLS fibers.}
\label{fig:range_detector}
\end{wrapfigure} 
This approach is taken since the expected energy resolution of traditional sampling calorimeters used in high energy physics may be inadequate due to poor shower statistics in this energy regime. Finally, the detector will be surrounded with a cosmic ray veto system.
The ESS is a pulsed neutron source, with proton pulse lengths of 2.86\,ms and a 14\,Hz repetition rate~\cite{ESS}. The pulse timing can be used for rejection of backgrounds from the spallation process. With long pulse lengths, the neutron beam will be essentially continuous, the baseline proposal for trigger and data acquisition is to use two trigger levels. The first level will include a continuous, 'self-triggered' readout of time-stamped, digitized pulse samples from the calorimeter and cosmic ray veto subsystems to a software-based trigger. Candidate events identified by the first trigger level will select matching TPC data frames to be read out for event reconstruction and storage.

\section{The NNBAR calorimeter prototype}

The detection conditions of the calorimeter are quite different from what is customary in high energy and nuclear physics. The main obstacles are kinetic energies of charged particles that are often too high for stopping by ionization only but too low for meaningful hadronic calorimetry. Gamma showers below the GeV region can only reach limited resolution, and the angle of incidence of gammas on the EM calorimeter will be unknown within a large interval. An environment with a large flux of fast and thermal neutrons may also pose additional consideration in the design. Detector simulations may not be well understood under these conditions and benchmarking simulations against beam test data of prototypes is a major goal. To this end, a calorimeter prototype is being constructed and tested over the next 2--3 years to validate the design for the first stage of the NNBAR experiment. The prototype will provide an R\&D test stand for studies of the response to the annihilation products: protons, pions (charged and neutral), and studies of the response to different backgrounds, particular to the spallation source environment. The following sections describe the mechanical setup and readout electronics for the Hadronic Range Detector (HRD) and the Lead glass EM Calorimeter (LEC). 

\subsection{Hadronic Range Detector}

\begin{figure}[H]
\begin{center}
\includegraphics[width=.9\linewidth]{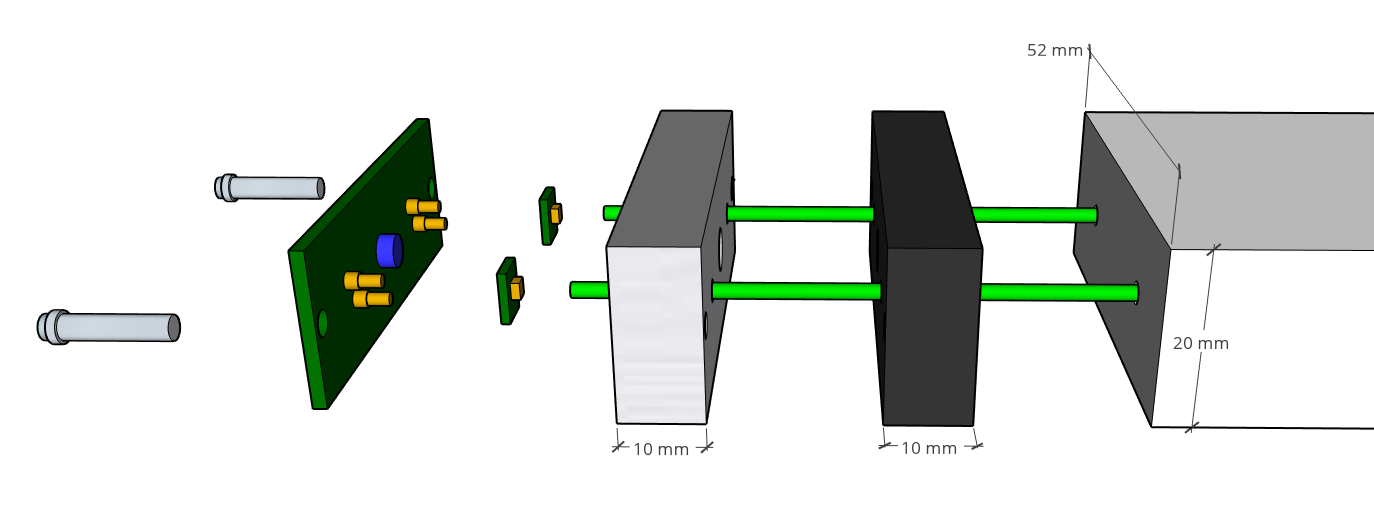}
\setlength{\belowcaptionskip}{-12pt}
\caption{CAD drawing of an exploded view of the scintillator stave readout components. The WLS shifting fibers are guided to the SiPMs with a plastic fiber guide bar and an electronics enclosure that positions the SiPMs precisely. The enclosure is connected by screws to the fiber guide bar. The fiber guide bar is glued to the scintillator stave.}
\label{fig:electronics}
\end{center}
\end{figure}

The HRD is composed of perpendicular layers of polystyrene plastic scintillator staves. The staves were extruded at the FNAL-NICADD Extrusion Line Facility~\cite{fermilab}. The dimensions are $5.18\times2\times50$\,cm$^{3}$, and the coating is TiO$_{2}$ which provides diffuse reflection, greatly localizing the scintillator photons for capture by the WLS fibers. Each stave has two 2mm extrusion holes for wavelength shifting (WLS) fibers of diameter 1.8\,mm. A CAD drawing of the prototype HRD is shown in Figure~\ref{fig:range_detector}. An exploded view of the electronics and mechanical setup are shown in Figure~\ref{fig:electronics}. The staves are assembled by first threading the WLS fibers through the extrusions. The WLS fibers are glued to plastic fiber guide bars on both ends of the stave. The fiber guide bars are glued to the scintillator staves. The WLS fibers are then coupled to silicon photomultipliers (SiPMs) mounted on a carrier board. The carrier boards are pressed flush to the WLS fibers and make electrical contact with a circuit board through spring loaded pins. The electronics manifold is secured with screws through the circuit board, SiPM holder, and fiber guide bar. The design of the enclosure the and carrier boards are inspired by the experiment~\cite{mu2e}. The circuit board includes an LED for calibration of the SiPM gain. The HV, LV, and SiPM signals are taken on- and off-detector through a twisted-pair ribbon cable. 

\begin{figure}
\centering
\begin{subfigure}[b]{.45\linewidth}
\includegraphics[width=\linewidth]{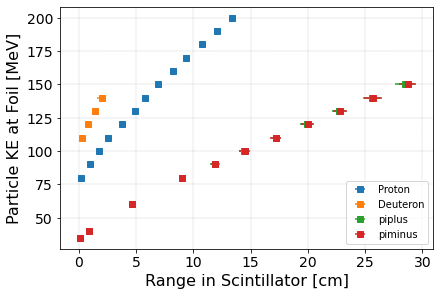}
\caption{}\label{fig:range}
\end{subfigure}
\begin{subfigure}[b]{.45\linewidth}
\includegraphics[width=\linewidth]{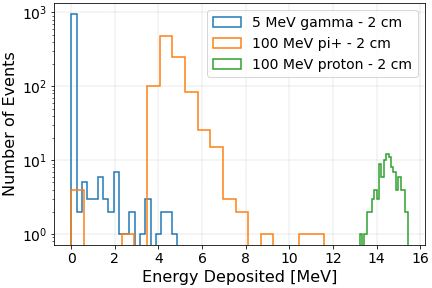}
\caption{}\label{fig:hists}
\end{subfigure}
\caption{Simulations of particle interactions in the HRD: (a) the range of particle in the scintillator plastic after passing through vacuum tube and other detector material (b) energy deposited in 2\,cm of scintillator plastic for different particles }
\label{fig:wls}
\end{figure}

GEANT-4~\cite{geant4} simulations have provided information on the energy resolution for various thicknesses of scintillators. Figure~\ref{fig:hists} shows simulations of the range in the scintillator plastic of particles with expected kinetic energies at the target foil. The chosen scintillator thickness of 2\,cm will induce energy deposits of close to 8\,MeV for charged pions, and close to 18\,MeV for protons. Figure~\ref{fig:hists} shows simulations of energy deposits in 2\,cm of plastic by nuclear gammas created by neutron capture in the carbon target along with energy deposits from $\pi^{+}$ and protons. A threshold can be used to remove the low energy gammas at the expense of efficiency. Additionally, these gammas will deposit all of their energy in a single scintillator stave, while pions will pass through a number of layers.

\subsection{Lead glass EM Calorimeter}

\begin{wrapfigure}[18]{r}{5cm}
\includegraphics[width=5cm]{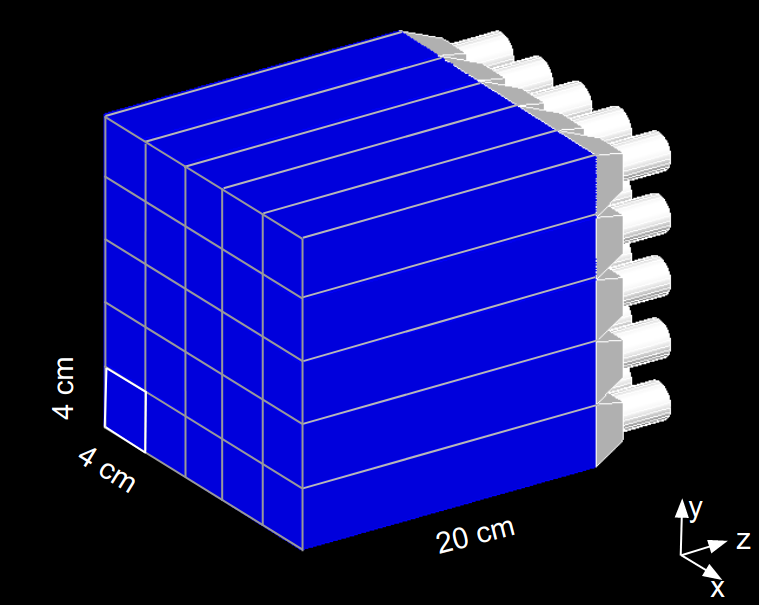}
\caption{A GEANT-4 simulation of the Lead glass EM Calorimeter. The detector is composed of $4\times4\times20$\,cm$^{3}$ lead glass blocks coupled to PMTs through plastic waveguides.}
\label{fig:my_label}
\end{wrapfigure}
\begin{figure}
\centering
\begin{subfigure}[b]{.48\linewidth}
\includegraphics[width=\linewidth]{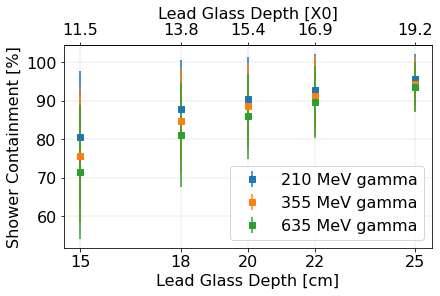}
\caption{}\label{fig:pb_depth}
\end{subfigure}
\begin{subfigure}[b]{.47\linewidth}
\includegraphics[width=\linewidth]{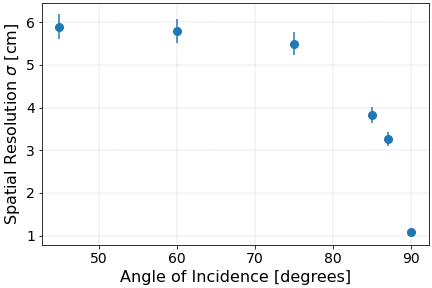}
\caption{}\label{fig:res}
\end{subfigure}
\caption{Simulations of optimization of the lead glass block dimensions: (a) particle shower containment in longitudinal direction as function of lead glass depth (b) spatial resolution of the point of impact of a 635\,MeV gamma onto a grid of 20 $4\times4$\,cm$^{2}$ lead glass blocks as a function of angle of incidence}
\label{fig:depth}
\end{figure}

\begin{figure}
\centering
\begin{subfigure}[b]{.24\linewidth}
\includegraphics[width=\linewidth]{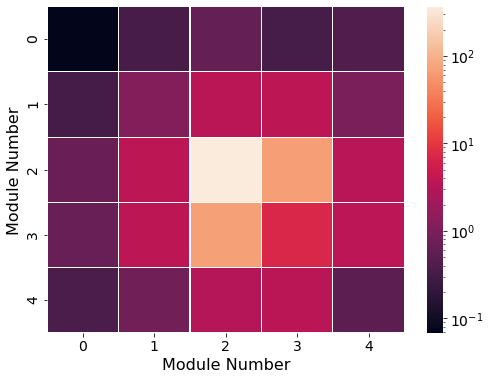}
\caption{}\label{fig:4_90}
\end{subfigure}
\begin{subfigure}[b]{.24\linewidth}
\includegraphics[width=\linewidth]{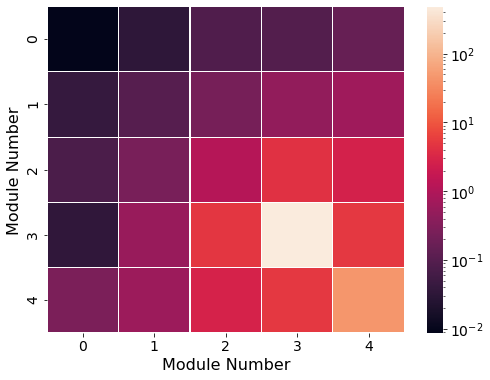}
\caption{}\label{fig:2_90}
\end{subfigure}
\begin{subfigure}[b]{.24\linewidth}
\includegraphics[width=\linewidth]{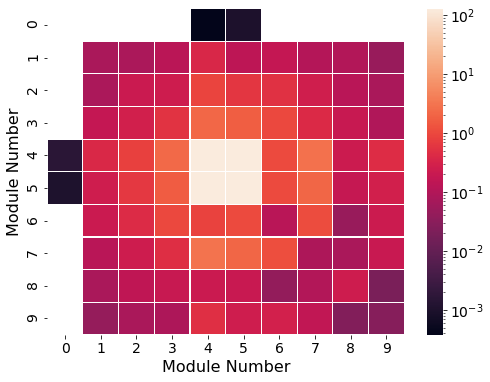}
\caption{}\label{fig:4_45}
\end{subfigure}
\begin{subfigure}[b]{.24\linewidth}
\includegraphics[width=\linewidth]{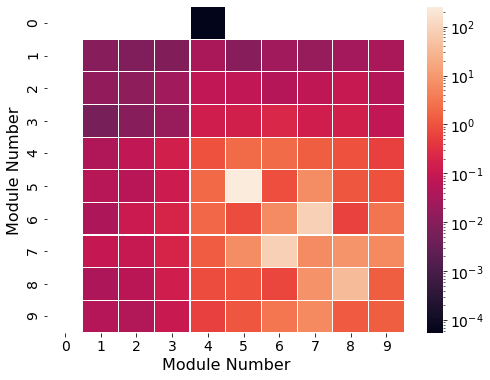}
\caption{}\label{fig:2_45}
\end{subfigure}
\caption{Simulation of energy deposited by a 635\,MeV gamma in lead glass blocks with varying block size and particle incidence angle: (a) $4\times4$\,cm$^{2}$, 90$^{\circ}$ incidence (b) $4\times4$\,cm$^{2}$, 45$^{\circ}$ incidence (c) $2\times2$\,cm$^{2}$, 90$^{\circ}$ incidence (d) $2\times2$\,cm$^{2}$, 45$^{\circ}$ incidence}
\label{fig:2D_blocks}
\end{figure}

 
The LEC is composed of $4\times4\times20$\,cm$^{3}$ ZF2 type lead glass blocks. The baseline readout is PMTs though other solutions are being explored including SiPM arrays. GEANT-4 simulations were done to determine the needed depth and granularity of the blocks. The energy resolution for $\pi^{+/-}$ is expected to be poor as the particles will undergo interactions with detector material before reaching the lead glass. The lead glass is then optimized to reconstruct gammas from the decay of $\pi^{0}$ at the foil by measuring the opening angle between the photons and the photon energies. Contrary to the standard use of EM calorimetry, the angle of incidence has a wide range. Figure~\ref{fig:pb_depth} shows simulations of the percent of the longitudinal containment of the shower produced by gammas with 210\,MeV, 355\,MeV, and 635\,MeV kinetic energy as a function of lead glass depth. 20\,cm was chosen for the prototype lead glass depth. The area of the lead glass blocks was chosen to be close to one Moli\`ere radius so that the shower energy is deposited among multiple blocks. The spatial resolution versus angle of incidence of the  grid of $4\times4$\,cm$^{2}$ blocks is shown in Figure~\ref{fig:res}. Figure~\ref{fig:2D_blocks} shows a simulation of the energy deposited by 635\,MeV gammas in a grid of lead glass blocks with varying block size and angle of incidence. Using the amplitude of the Cherenkov signal, the barycenter of the shower in the lead glass blocks can be computed. The resolution reported is the mean of the residuals between the true point of impact and the reconstructed point.  

\section{Summary} 

The HIBEAM/NNBAR experimental program at the ESS is a two-stage experiment to search for transitions of neutrons to sterile neutrons and antineutrons. The calorimetery requirements are challenging as the detector must reconstruct the invariant mass of two nucleons using information from particles with kinetic energies less than 600\,MeV. A prototype calorimeter for the HIBEAM/NNBAR experimental program is under construction. The prototype response will provide input to full-detector simulations and validate the energy and spatial resolution of the calorimeter, and is planned to be installed at the ESS testbeam line for detector validation and studies of backgrounds specific to the ESS spallation process.

\section{Acknowledgments}
The authors gratefully acknowledge project grant support from Vetenskapsr\aa det.

\section*{References}

\bibliography{TIPP2021}

\end{document}